# Segmentized quarantine policy for managing a tradeoff between containment of infectious disease and social cost of quarantine

**Jungwoo Kim[a], Taesik Lee[a]**

*[a] KAIST, Daejeon, South Korea*

## Abstract

By the end of 2021, COVID-19 had spread to over 230 countries, with more than 5.4 million deaths. To contain the disease spread, many countries have deployed non-pharmaceutical intervention strategies, most notably contact tracing and self-quarantine policy. We have observed that containment of disease spread by such social distancing policy come at a large social cost, and prolonged pandemic raised the necessity of more sustainable policy with the least disruption to economic and societal activities. This research aims to investigate a segmentized quarantine policy where we apply different quarantine policies for different population segments with a goal of better managing the tradeoff between the benefit and cost of a quarantine strategy. Motivation for a segmentized policy is that different population groups, e.g., school students vs adults with jobs, exhibit different patterns of societal activities, thereby imposing different risks to disease spread. We define a segmentized quarantine policy in two dimensions – range of contact tracing and quarantine period, and determine the two parameters for each population segment to achieve two objectives: to minimize the total number of infected cases and to minimize the total days of self-quarantine. We use Agent-based Epidemics Simulation to evaluate the quarantine policies, and Evolutionary Algorithm is used to obtain the Pareto front of our problem. Our results demonstrate the effectiveness of the segmentized quarantine policies, and we identify the conditions where they outperform the uniform policy. We also find in the Pareto optimal solutions that only some population segments are recommended special policy features while other segments are subject to the conventional policy. The results suggest that segmentized quarantine policy is valid in terms of efficiency and sustainability, and the suggestions and framework presented are expected to be of great help in establishing public health decisions to prepare for an upcoming pandemic like COVID-19.

## 1. Introduction

COVID-19 has made a huge impact on our lives, and the continual emergence of variants such as delta and omicron are bringing new waves of uncertainty and fear. Although the vaccination rate is more than half worldwide, hospitals are still full of severe patients and suffer shortage of resources [1]. One of the main strategies applied around the world is contact tracing and self-quarantine. According to CDC guidelines, asymptomatic people who are designated as close contacts are self-quarantined and be monitored for 14 days after their last exposure [2]. The purpose is to break the chains of transmission through quickly identifying and isolating potential cases, who have been exposed to the diseases and going through an incubation period [3]. This intervention is estimated to have a significant effect on reducing the reproduction number of COVID-19 [4]. However, entering the third year of the outbreak, damages caused by this strong policy is becoming a grave concern as psychological, social, financial losses are reported in many countries [5]-[8]. Prolonged pandemic also raises the concern of pandemic fatigue, which leads to reduction of policy compliance [9]. These phenomena are raising the necessity of a sustainable and effective quarantine policy against COVID-19 and another potential upcoming pandemic.

The intensity of a quarantine policy is mainly determined by two parameters: the range of contact tracing and the quarantine period. Range of contact tracing concerns how broadly we designate those who need to be quarantined. The CDC standard for close contact is "someone who was within 2 meters of an infected person for at least 15 minutes within 24 hours starting from 2 days before illness onset until the time the patient is isolated" [2]. This may include household members, colleagues in the workplace, or friends in school. Increasing the range of contact tracing means designating more people to be in such close contact by broadening the standard. This would lower the reproduction number and thus reduce the number of upcoming cases by isolating more unreported cases from the community. Conversely, reducing the range of contact tracing has the effect of accelerating the disease spread. A quarantine period is related to the incubation period of the virus. People become symptomatic and infectious at different rates, and about 97% of people show symptoms within 11 days [10]. Therefore, by increasing the quarantine period, it is possible to isolate more unreported cases among close contacts and suppress the spread of disease.

If we are only concerned with containing the spread of disease, a quarantine policy with higher intensity – broader contact tracing and longer quarantine period – will be an obvious decision. Unfortunately, though, executing a quarantine policy comes at a social cost, and increasing its intensity causes more incidental quarantine-related costs. In this respect, a tradeoff arises between the effectiveness of disease containment and the cost born by the society at large, and this tradeoff must be carefully managed. The main purpose of this research is to investigate the effectiveness of a segmented quarantine policy where we apply different quarantine policies for different population segments with a goal of better managing the tradeoff between the benefit and cost of a quarantine strategy. Motivation for a segmented policy is that different population groups, e.g., school students vs adults with jobs, exhibit different patterns of societal activities, thereby imposing different risks to disease spread. Features such as an individual's daily workplace or age affect the number and type of people they contact, and the susceptibility for disease. Because of these differences, a uniform quarantine policy may be too conservative for some and not strict enough for others. As an alternative, applying heterogeneous quarantine policies, appropriately designed for different population segments, can reduce unnecessary quarantine costs while achieving the same level of disease containment. Using an agent-based epidemics simulation model and multi-objective optimization, we demonstrate that a segmented quarantine policy is potentially an effective tool to manage the tradeoff in quarantine policy design.

## 2. Related Research

Several studies have been conducted to estimate the effectiveness of policies related to COVID-19. Amin et al. used a multi-objective genetic algorithm that proposes strategies considering its economic consequences [11]. They used a deterministic compartmental model to measure the intervention settings. Transition parameters such as diagnose rate are used as a decision variable, and also one of the cost functions. Igor et al. presented a framework for optimizing the required levels of public health policy, coupled with a nonlinear model [12]. They also extended and validated a compartmental mathematical model to evaluate the effects of interventions. Ma et al. optimized the non-uniform lockdown policy for pandemic control [13]. They built a SIR based network model to optimize the lockdown rate for each county in New York city. Acemoglu et al. optimized targeted lockdowns with a multi-group SIR model [14]. They found that optimal policies differentially targeting risk/age groups significantly outperform optimal uniform policies. Most of the preceding studies used the compartmental SD model for policy evaluation, which has the disadvantage of not reflecting the heterogeneity of the members of community. Also, their decision variables or cost indicators are not intuitive enough to be a direct advice to the decision makers. In addition, there are studies dealing with the application of non-uniform lockdown policies to members, but studies on the intensity of contact tracing self-quarantine policies, which are more relaxed than lockdown, have not been conducted. Therefore, this study aims to evaluate the effectiveness of the non-uniform quarantine policy and obtain recommendations for sustainable policy establishment using direct and intuitive indicators and decision variables.

## 3. Problem Definition and Methodology
### 3.1. Problem Statement
The objective of this research is to propose a more sustainable policy that considers both disease containment and social cost. To this end, we would like to analyze the efficiency and validity of the segmented policy by comparing the optimized segmented policy and uniform policy. We define a segmented quarantine policy in two dimensions – range of contact tracing and quarantine period, and determine the two parameters for each population segment to achieve two objectives: to minimize the total number of infected cases and to minimize the total days of self-quarantine. When applying the segmented policy, the population will be classified into 10 segments according to their age group and daily workplace, and different quarantine policies will be applied. We seek to find policy suggestions such as the conditions of which segmented policy outperforms the uniform policy and recommended policy intensities for each segment, by analyzing the Pareto front and optimal solutions of both policies.

### 3.2. Agent-based epidemics simulation
An agent-based epidemic simulation was built to evaluate each policy settings. 10,000 agents were sampled from the census data of Seoul, Korea [15]. The age group, family composition, and daily workplace of each individual were set to reflect the heterogeneity of the population. **Figure 1** shows the state transition graph of the simulation. All the agents are initially in the susceptible state, and only one agent is designated as an initial seed, set as a nonsymptomatic-infected state, and serves as a starting point of an epidemic. The transmission of the disease is caused by the contact between agents. The contact mechanism between agents is fabricated based on the FluTE model developed by Dennis et al. [16]. Each agent has three fields of daily activity, and contact occurs independently in each contact field. The contact fields consist of *Community*, *Neighborhood* and *Spot* each reflecting various scale of societal activities. The daily workplaces, which segmentizes the agents, is a daytime *spot* which includes *household, playgroup, daycare, elementary school, middle school, high school,* and *workplace.* The family members visit the same nighttime spot every day, which reflects the pattern of returning home at the end of the day. Each pair of agents in the same contact field make contact by the same contact probability every day. The contact probability is set differently according to the type of contact field, and age types of the agents, which follows the FluTE's.

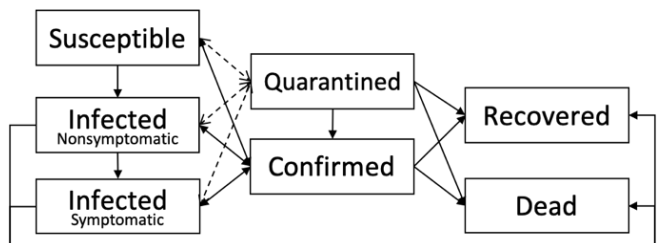

**Figure 1:** State transition graph of Agent-based epidemics simulation

After making contacts in each contact fields, agent who made contact with an infectious agent is considered as *close contact*. With certain probability, they turn into *nonsymptomatic-infected*. This probability is calculated by the product of a constant called *p_trans* and *age dependent susceptibility*. Age dependent susceptibility increases with age, and p_trans has been calibrated to suit the reproduction number of South Korea [17, 18]. When an agent is infected, the periods between the states such as the incubation period and recovery period are decided according to data-driven probability distribution [19]-[21]. Whether the *close contact* needs to be quarantined is determined by the range of contact tracing. If it is 1, all *close contacts* are ordered to be quarantined. And if it is less than 1, only some of them are ordered. If greater than 1, other agents within the same contact fields are additionally ordered. Those ordered ones start their quarantine from the moment their infector gets confirmed until the quarantine period ends.

### 3.3. Multi-objective Evolutionary Algorithm
To obtain the Pareto front of segmentized quarantine policy, *Evolutionary Algorithm* is used [23]. EA is a modified genetic algorithm that is applicable to a multi-objective optimization program. In addition to a repository called *Population* in general GA, there is another one called *Archive* which is a set of Pareto-dominating solutions. One generation of EA is executed as shown in **(a)** of **Figure 2**. In the first step, the parents from the old population and the old archive are selected according to their fitness value to produce offspring. In the second step, these offspring are evaluated by our simulation to obtain cost function values. In the third step, the new archive is updated by checking the Pareto-dominance between the newly created offspring and old archive members. The offspring is produced by five operators in **(b)** of **Figure 2**. Crossover is done by selecting one parent from the archive and another from the population. For archive mutation and population mutation, the 4-point mutation is done to the parents from each repository. *Dominance count* is used as the fitness value when selecting a parent from the population and archive. This helps to uniformly obtain the points forming the Pareto front, and fill in the sparse section.

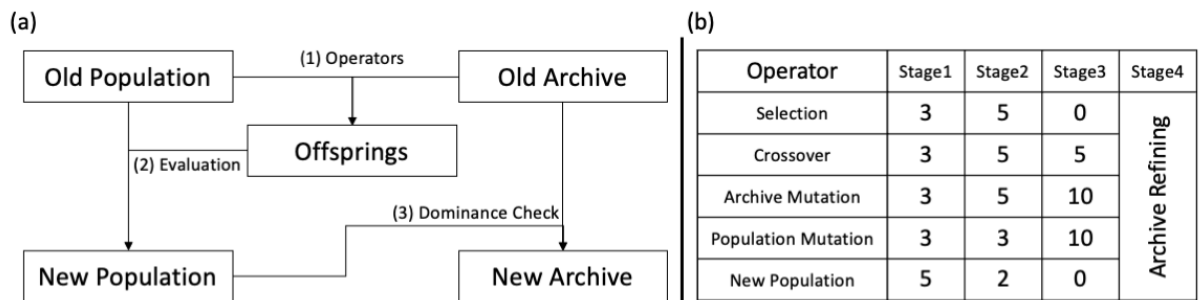

**Figure 2: (a)** One generation of Evolutionary Algorithm, **(b)** Operator ratio of Evolutionary Algorithm in each stage

## 4. Results and Discussion
### 4.1. Experimental setting
The population of the epidemic simulation is 10,000, with the same population composition and initial seed for every evaluation. Since the agent-based epidemics simulation has some uncertainty, Sample Average Approach (SAA) was utilized [22], which simply repeats the simulation several times to use the average as the estimated value of the cost function. Thus, the simulation was repeated 10 times in every evaluation. The range of contact tracing is set between 0 and 3.0, and the quarantine period is set from 0 to 21 days. When executing the evolutionary algorithm to optimize the segmentized policy, the number of chromosomes in a generation was set to 100, and evolution is performed in four stages as shown in **(b)** of **Figure 2**. The numbers in the table are the proportions of offspring made by each operator on each stage. 100 generations were executed for each stage. In the last stage, solutions in the archive are reevaluated and updated repeatedly to remove the outliers. The evaluation of the uniform quarantine policy was done with grid search rather than EA since the number of combinations was not too large.

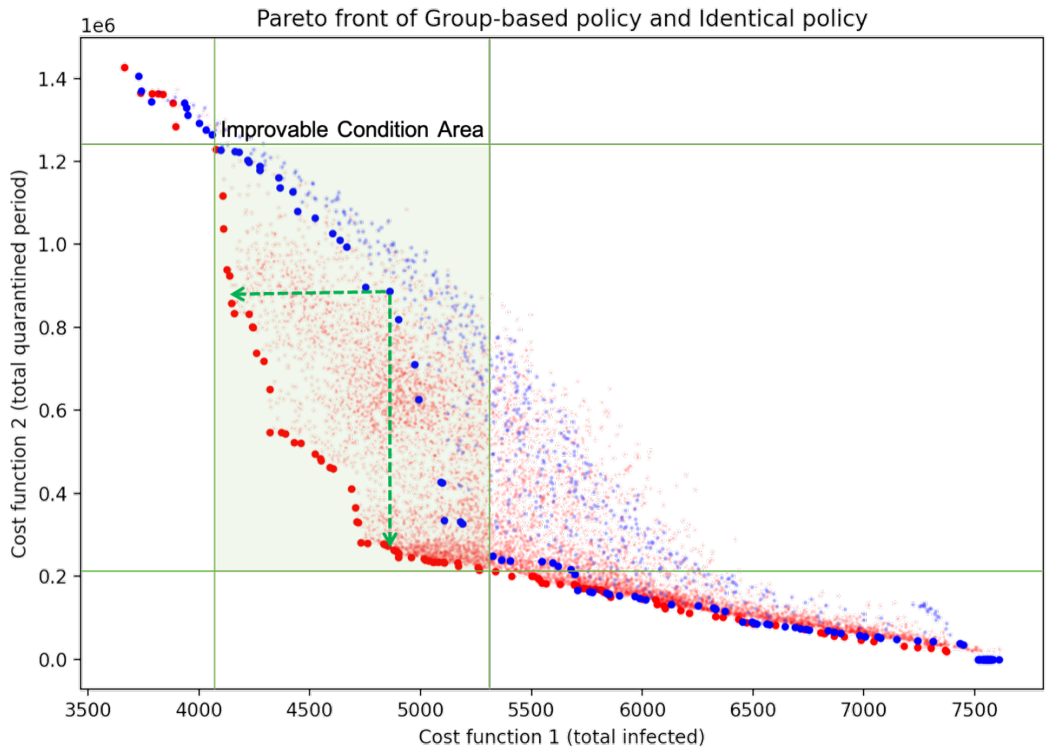

**Figure 3:** Pareto front of Segmentized policy and Uniform policy. Red dots are solutions of segmentized policy, and blue dots are solutions of the uniform policy. Large dots are dominating solutions forming the archives, which also are Pareto-optimal solutions of each policy. The area colored in green represents the *Improvable condition area* in which the segmentized policy outperforms the uniform policy. Green arrows indicate the cases where segmentized condition can reduce one cost function the most without losing another cost function.

### 4.2. Results
**Figure 3** shows the Pareto front of both segmentized and uniform policy. The part of the graph where the difference occurs between the red and blue large dots is where the segmentized policy outperforms the uniform policy. And we call this *Improvable condition*. Since applying different policies for each segment requires difficult social agreement and expensive implementation costs, a uniform policy is proposed outside this area. But within this area, an optimized segmentized policy can reduce total infection cases up to 13.7% with same social cost. It can also reduce the total quarantine period up to 68.8% without any loss of disease containment (green arrows in **Figure 3**). Thus, it is a more efficient strategy that can reduce both the transmissions and the social cost.

### 4.3. Discussion
First of all, the Pareto-optimal solutions of the uniform policy (i.e., large blue dots in **Figure 3**) are analyzed. The optimized uniform policies have a high range of contact tracing in common (avg: 2.83, std: 0.036). That is, when applying the uniform policy, it is recommended to quarantine as many people as possible, no matter what the importance of cost functions are. Also, the quarantine periods of those solutions have a significant correlation with cost functions (corr with cf1: -0.976, corr with cf2: 0.986). Thus, the quarantine period acts as a key parameter to decide which of the Pareto-optimal solutions to choose.

**Table 1** shows the segmentized Pareto-optimal solutions in the improvable condition (i.e., large red dots in **Figure 3**). The tendency is particularly significant in several segments. Segment 4-6 consist of young agents attending schools. The solutions have a moderate-level of both decision variables in common. In other words, same level of policy is recommended constantly no matter which Pareto-optimal point is selected. The policy levels to these segments do not need to change as the importance of the cost functions changes. It seems that this is because the population of the segments is small, but the spot sizes of schools are large. And most families with those young agents also include adults. Therefore, they play a big role in the spread of disease. Segment 8 consists of adults going to the workplaces in the daytime. The ranges of contact tracing of this segment are high in all the Pareto-optimal solutions. Also, the quarantine periods have a significant negative correlation with the cost function 1. Therefore, the broadest range of contact tracing is recommended in any circumstances. Also, the quarantine period of this segment acts as a key parameter to decide which of the Pareto-optimal solutions to choose. This is probably due to the large population of this segment, and most families include a member of this segment. Finally, segment 9 consists of aged people who stay at home during the day. The Pareto-optimal quarantine period of this segment was near 0, which means the quarantine policy is not recommended to this segment in any case as it does not help suppressing the disease spread but rather only increases social cost. This is because their social activities are small, so the risk as a source of infection in the community is very low. Also, the quarantine takes place at home, and they often form a family among them, so there is eventually no effect of self-quarantine. Other segments have relatively high variance and low correlation with the decision variables. This implies that they have no significant tendency within the Pareto-optimal solutions, so the conventional level of policy is recommended to these segments to minimize the implementation cost and difficulty of social agreement.

**Table 1:** Segment-wise statistics of Pareto-optimal solutions in the improvable condition. Average, variance, correlation with cost functions 1 and 2 are calculated for each decision variable respectively.

| | | Population Segments | | | | | | | | | |
|---|---|---|---|---|---|---|---|---|---|---|---|
| | | 1 | 2 | 3 | 4 | 5 | 6 | 7 | 8 | 9 | 10 |
| **Member property** | | young, household | young, playgroup | young, daycare | young, elementary | young, middle | young, high school | adult, household | adult, workplace | old, household | old, workplace |
| **Range of Contact tracing (Decision variable 1)** | Average | 1.704 | 1.554 | 1.618 | **1.642** | 2.362 | 1.676 | 2.448 | **2.840** | 1.506 | 2.164 |
| | Variance | 1.001 | 0.556 | 0.747 | **0.074** | **0.156** | **0.154** | 0.436 | **0.036** | 1.139 | 1.077 |
| | Corr w/ CF1 | 0.064 | 0.251 | -0.078 | -0.187 | 0.063 | -0.096 | 0.049 | -0.171 | 0.054 | 0.045 |
| | Corr w/ CF2 | 0.067 | -0.257 | 0.112 | 0.142 | -0.030 | 0.032 | -0.083 | 0.137 | -0.148 | -0.019 |
| **Quarantine period (Decision variable 2)** | Average | 4.760 | 9.820 | 8.320 | **15.300** | 6.040 | 4.920 | 12.320 | 10.280 | **0.620** | 4.320 |
| | Variance | 31.615 | 38.967 | 30.181 | **15.153** | **5.060** | **3.830** | 39.242 | 66.940 | **4.404** | 38.467 |
| | Corr w/ CF1 | -0.100 | -0.179 | -0.195 | -0.269 | 0.151 | -0.058 | 0.163 | **-0.902** | -0.150 | -0.287 |
| | Corr w/ CF2 | 0.138 | 0.201 | 0.094 | 0.346 | -0.136 | 0.056 | -0.131 | **0.891** | 0.046 | 0.307 |
| **Population** | | **345** | **69** | **164** | **494** | **273** | **320** | **1958** | **5083** | **1087** | **207** |

## 5. Conclusions

In this study, both segmentized and uniform quarantine policies are optimized using an agent-based epidemics simulation and evolutionary algorithm. The Pareto front of both policies show that there are improvable conditions where segmentized policies significantly outperform the uniform policy. Also, guidelines for determining sustainable quarantine-related parameters are presented through analysis of Pareto-optimal solutions. Range of contact tracing did not change no matter which Pareto-optimal solution was selected. When uniform policy is to be applied, the highest range of contact tracing is recommended. When applying a segmentized policy, special level of contact tracing is recommended for only some segments. The choice of solution in the Pareto front may vary depending on the relative importance of benefit and cost of a quarantine strategy, or limitation of the cost functions (e.g., shortage of medical resources). And the key parameter to manage the tradeoff is the quarantine period. In particular, in the segmentized policy, the quarantine period of segment 8 served as a key parameter for selecting from the Pareto-optimal solutions. Although the suggestions for policy decisions presented in this study may vary slightly depending on the characteristics of the community, segmentized policies optimized for various communities can be applied using the framework proposed in the study. These findings are expected to be of great help in establishing sustainable public health decisions to prepare for an upcoming pandemic like COVID-19.

## Acknowledgement
I wish to extend my special thanks to Hyunjin Lee, Hyelim Shin, Hyewon Park, and Junyoung Jung who helped develop the agent-based epidemics simulations used in the research.